\documentclass[12pt]{article}
\usepackage{amssymb}
\usepackage{graphicx,psfrag, subfigure}
\usepackage{amsmath}
\usepackage{bbold}
\usepackage{blkarray}
\usepackage{dsfont}
\usepackage{yfonts}
\usepackage{setspace}
\usepackage{soul}
\usepackage{mathrsfs}
\numberwithin{equation}{section}
 \usepackage{relsize}
\usepackage{jheppub}
\usepackage[mathscr]{eucal}
\usepackage{mathtools} 
\usepackage{centernot}

\def\ms{\mathsf}

\def\be{\begin{equation}}
\def\ee{\end{equation}}
\def\bea{\begin{eqnarray}}
\def\eea{\end{eqnarray}}

\def\be{\begin{equation}}
\def\ee{\end{equation}}
\def\bea{\begin{eqnarray}}
\def\eea{\end{eqnarray}}

\usepackage{lmodern}

\usepackage[T1]{fontenc}
\usepackage[utf8]{inputenc}

\newcommand{\pfrac}[2]{\ensuremath{{\frac{\partial #1}{\partial #2}}}}

\newcommand{\td}{\ensuremath{{\text{d}}}}

\def\sqrtexplained#1{
  \begingroup
    \sbox0{$#1$}
    \def\underbrace##1_##2{##1}
    \sbox2{$#1$}
    \dimen0=\wd0 \advance\dimen0-\wd2
    \mathrlap{\sqrt{\phantom{\displaystyle#1}\kern\dimen0 }}
    \hphantom{\sqrt{\vphantom{\displaystyle#1}}}
  \endgroup
  #1}

\begin{document}

\begin{titlepage}

\setcounter{page}{0} \baselineskip=15.5pt \thispagestyle{empty}

\bigskip\
\begin{center}

{\Large \bf 
Action Principle for Isotropic General Relativity}\vspace{10pt}

\vskip 5pt
\vskip 15pt
\end{center}
\vspace{0.5cm}
\begin{center}
{Thomas C. Bachlechner}

\end{center}\vspace{0.05cm}

\begin{center}
\vskip 4pt
\textsl{Department of Physics, University of California, San Diego\\La Jolla, CA, 92093}
\end{center}

{\small  \noindent  \\[0.2cm]
\noindent
We study the generally covariant theory governing an isotropic spacetime  region ${\cal M}$ with uniform energy density. Gibbons, Hawking and York showed that fixing the induced boundary metric yields a well-posed variational problem. However, as we  demonstrate, fixing the boundary metric violates general covariance and allows the  mass of a back hole to vary. This observation has dramatic consequences for path integrals: A sum over  spacetimes with fixed boundary metrics is a sum over classically distinct black holes. Instead, we merely demand that coordinates exist such that the metric at the boundary $\partial \cal M$ is the Schwarzschild-(A)dS metric of fixed mass ${\ms M}$ and two-sphere radius $\ms R$.  We derive the action that yields a well-posed variational problem for these physical boundary conditions, $\delta \ms M|_{{\partial {\cal M}}}=\delta \ms R|_{{\partial {\cal M}}}=0$. The action vanishes for all stationary and isotropic spacetimes. A vanishing action implies that both a Schwarzschild black hole and pure de Sitter space each have one unique semiclassical state. Our results provide a novel and radically conservative approach to several long-standing issues in quantum gravity, such as the wavefunction of the universe, the black hole information paradox, vacuum decay rates and the measure problem of eternal inflation.

}

\vspace{0.3cm}

\vspace{0.6cm}

\vfil

\begin{flushleft}
\small \today
\end{flushleft}
\end{titlepage}
\tableofcontents
\newpage

\section{Introduction}\label{intro}
It is generally accepted that the prevailing understanding of gravity and (semiclassical) quantum theory are  incompatible. This conflict is manifest in several failures of the theory, like the black hole information paradox \cite{Hawking:1976ra}, the measure problem \cite{boltzmann1895certain,albrecht_2004,susskind_2007,Carroll:2005ah,Aguirre:2011ac,Rubakov:1988jf,Hartle:1983ai,Linde:1983mx,Vilenkin:1984wp,Vilenkin:1986cy,Dyson:2002pf}, or the prediction that black holes  densely populate spacetime \cite{Fischler:1988ia,fmp2}. No satisfactory resolution of these serious problems is known. The usual approach to semiclassical gravity relies on self-consistent but unphysical boundary conditions fixing the induced spacetime metric on a coordinate-hypersurface in nature, where the principle of general covariance is broken to covariance under boundary-preserving diffeomorphisms\footnote{The term ``unphysical'' specifically refers to the fact that no observer can determine the boundary conditions: the induced metric is physically unobservable. For example, the metric on a global spatial slice would contain information about causally disconnected spacetime regions, which immediately leads to a measure problem and spoils predictions. A theory with an unresolved measure problem may be philosophically entertaining, but is ultimately inadequate.}. In this work we will see that  isotropic general relativity\footnote{In contrast to ``general relativity'', the  gravitational theory defined by a fixed induced boundary metric could more accurately be referred to as  ``boundary-preserving relativity'', since the theory is only covariant under boundary-preserving diffeomorphisms. In this work the term ``covariant'' refers to covariance under general diffeomorphisms, unless noted otherwise.} has a  semiclassical description that may resolve many of the pathologies that arise when fixing the induced metric on the boundary.

Einstein  formulated the general theory of relativity with the requirement that the laws of nature be generally covariant, i.e. covariant with respect to arbitrary (smooth) coordinate transformations \cite{Einstein:1916vd}. The laws of classical physical systems consist of the equations of motion and boundary conditions that determine the solutions to the equations of motion.  The vacuum equations of motion are generally covariant,
\be\label{einstein}
G_{\mu\nu}=-8\pi G \rho g_{\mu\nu}\,,
\ee
where $G_{\mu\nu}$ is the Einstein tensor, $g_{\mu\nu}$ is the spacetime metric and $\rho$ is the vacuum energy density. The boundary conditions, on the other hand, are often difficult to express in a  covariant manner. In this work we are radically conservative and demand general covariance. Correspondingly, we regard boundary conditions as physical only if they are covariant under all smooth coordinate transformations. Requiring general covariance does not affect the equations of motion (\ref{einstein}). However, requiring general covariance affects whether the equations of motion arise from a variational principle $\delta S=0$. Boundary conditions for coordinate dependent variables (like the induced metric) are meaningless unless we break covariance and fix the coordinates.  
Boundary conditions for covariant variables, on the other hand, are meaningful and yield a well-posed variational problem even if the coordinates at the boundary are allowed to vary. Consequently, general relativity allows only boundary conditions for generally covariant variables.

It has long been known that some actions for gravity do not give  rise to equations of motion \cite{Arnowitt:1962hi,York:1972sj,Regge:1974zd,Gibbons:1976ue,Brown:1992br,Brown:1992bq}. For example, the Einstein-Hilbert action
\be
S_\text{EH}={1\over 16\pi G}\int_{\cal M}\td ^4 x\, \sqrt{g}{\cal R}\,,
\ee
contains second derivatives of the metric, and therefore does not have to a well-defined variational problem when fixing the metric components on the boundary of ${ \cal M}$. We see this explicitly from the variation
\be\label{ehvariation}
16\pi G\,\delta S_\text{EH}=\int_{\cal {\partial \cal M}} (d^3 x)_\rho ~\sqrt{g} \left(g^{\mu\nu} \delta^{\rho}_{\,\lambda}-g^{\mu\rho} \delta^{\nu}_{\,\lambda}\right)\delta \Gamma^\lambda_{\mu\nu}-\int_{\cal M} d^4 x ~\sqrt{g} \left(R_{\mu\nu}-{1\over 2}R g_{\mu\nu} \right)\delta g_{\mu\nu}\,,
\ee
where $\Gamma^\lambda_{\mu\nu}$ denotes the Christoffel symbols. The  principle of stationary action $\delta S_\text{EH}=0$ does not coincide with Einstein's equations (\ref{einstein}) when imposing  Dirichlet boundary conditions  $\delta g_{\mu\nu}|_{\partial \cal M}=0$. The problematic first term in the variation (\ref{ehvariation}) is a surface integral that can be eliminated by the inclusion of an appropriate boundary term \cite{,York:1972sj,Gibbons:1976ue},
\be
S_\text{GHY}=-{1\over 8\pi G} \sum_A \epsilon_A \int_{{\partial \cal M}_A}d^3 x~ \sqrt{\gamma} {\cal K}\,,
\ee
where $\epsilon=\pm 1$ is positive (negative) for a space-like (time-like) boundary ${\partial \cal M}_A$ with induced metric $\gamma_{\mu\nu}$ and ${\cal K}$ is the trace of the extrinsic curvature tensor. The boundary contributions to the variation of the combined action now become
\be\label{btss}
16\pi G\,\delta (S_\text{EH}+S_\text{GHY})|_{\partial {\cal M}}=- \sum_A \epsilon_A \int_{{\partial \cal M}_A}d^3 x~ \sqrt{\gamma} \left({\cal K}\gamma^{\mu\nu}-{\cal K}^{\mu\nu}\right)\delta \gamma_{\mu\nu}\,,
\ee
and thus vanish when demanding Dirichlet boundary conditions for the induced metric, $\delta \gamma_{\mu\nu}=0$. The action gives rise to a well-defined action principle, but unfortunately the boundary metric $\gamma_{\mu\nu}$ depends on the coordinate choice. If we insist on general covariance, the coordinates (and the associated part of the induced metric $\gamma_{\mu\nu}$) are allowed to vary freely, giving $\delta \gamma_{\mu\nu}\ne 0$, and thus non-vanishing boundary terms (\ref{btss}). We conclude that the variational principle $\delta (S_\text{EH}+S_\text{GHY})=0$ does not yield Einstein's equations if we impose generally covariant boundary conditions.

 In this work we will consider configurations with manifestly generally covariant boundary conditions. We demand that the spacetime at the boundary $\partial \cal M$ be related by a general diffeomorphism to the Schwarzschild-(anti) de Sitter metric,
\be\label{staticmetrici}
		\td s^2= -A({\ms  M},{\ms  R}) \td {\ms  T}^2 + A^{-1}({\ms  M},{\ms  R}) \td {\ms  R}^2 + {\ms  R}^2 \td\Omega_2^2\,,~~~A=1-{2G{\ms  M}\over{\ms  R}}-{8\pi G\rho\over 3} {\ms  R}^2,
\ee
where $\rho$ is the energy density and $\ms M$ is the black hole mass. The mass $\ms M$ depends on derivatives of the metric that remain unfixed if we merely fix the induced metric, $\delta \gamma_{\mu\nu}=0$. Therefore, fixing the induced metric does not fix the black hole mass. This observation  has consequences for path integrals in general relativity: instead of summing over black hole configurations with fixed mass, the path integral with fixed boundary metric would include a sum over black holes with distinct masses. This might give the impression that a black hole has a large number of microstates. Instead, the boundary conditions simply failed to fix the asymptotic mass and we inadvertently summed over an ensemble of black holes with varying masses. A similar statement holds for de Sitter spacetimes. We will find that when assuming general covariance, the action for static vacuum configurations vanishes, suggesting that Schwarzschild black holes of fixed mass and de Sitter space each have one semiclassical microstate.

Consider the most general spherically symmetric metric written in ADM form \cite{Arnowitt:1962hi},
\be
\label{eq:NLRMetricintro}
ds^2 = -{N_t^2}(t,r) \td t^2 + \Lambda^2(t,r) (\td r + {N_r} (t,r)\td t)^2 + R^2(t,r) \td \Omega_2^2 \,.
\ee
The lapse $N_t$ and the shift $N_r$ are arbitrary non-dynamical functions that set the gauge, while the variables $\Lambda$ and $R$ are dynamical. The form of the metric (\ref{eq:NLRMetricintro}) shows that the two-sphere radius $R$ is covariant under coordinate transformations, while $\Lambda$ is not covariant\footnote{We give the explicit transformation properties in Appendix \ref{app1}.}
\be\label{admrepss}
\{t,r\}\rightarrow \{\tilde{t},\tilde{r}\}\,:~~~R(t,r)\rightarrow R(\tilde{t},\tilde{r})\,,~~\Lambda(t,r)\nrightarrow \Lambda(\tilde{t},\tilde{r})\,.
\ee
The ADM variable $\Lambda$ transforms like a spatial scalar density and therefore is meaningless without a fixed coordinate choice. 
 
In order to formulate our physical boundary conditions (\ref{staticmetrici}), we use the mass ${\ms  M}$ and the two-sphere radius ${\ms  R}$ as variables. This change of variables was first proposed by Kucha\v{r} in \cite{Kuchar:1994zk}. We will follow much of their treatment but assume different boundary conditions. The metric then is given in terms of the new variables, the  momentum  $\pi_{\ms M}= -\ms T'$ conjugate to $\ms M$, and new (non-dynamical) Lagrange multipliers $\ms N_{t}$ and $\ms N_{r}$ as
\be
ds^2=\left(A^{-1}{R'^2}-A\pi^{2}_{\ms M}\right) \left[(\td r+ \ms N_r\td t)^2- \ms N_t^2 \td t^2\right]+ \ms R^2 \td \Omega_2^2\,,
\ee
where primes and dots denote $r$ and $t$ derivatives, respectively. In contrast to non-covariant transformation properties of the ADM variables in (\ref{admrepss}), both $\ms R$ and $\ms M$ transform covariantly,
\be
\{t,r\}\rightarrow \{\tilde{t},\tilde{r}\}\,:~~~\ms R(t,r)\rightarrow \ms R(\tilde{t},\tilde{r})\,,~~\ms M(t,r)\rightarrow \ms M(\tilde{t},\tilde{r})\,.
\ee
With a  suitable choice of boundary terms we find the gravitational action
\be\label{action1}
S_\text{G}=\int_{\cal M} \td t \td r~{\cal L}_\text{G}=\int_{\cal M} \td t \td r~\frac{{\ms  M}' \left[\left(\ms N_r^2-\ms N_t^2\right) {\ms  R}'-\ms N_r \dot{{\ms  R}}\right]+\dot{{\ms  M}} \left[\dot{{\ms  R}}-\ms N_r {\ms  R}'\right]}{A({\ms  M},{\ms  R}) \ms N_t}\,.
\ee
The Lagrangian density transforms like a scalar density of weight one, contains no derivatives of the lapse and the shift, and correspondingly we require no boundary conditions for these unphysical variables. Instead, Dirichlet boundary conditions for the mass and the two-sphere radius alone yield a well-defined variational principle,
\be\label{physicalbc}
\delta{\ms  M}|_{\partial \cal M}=\delta{\ms  R}|_{\partial \cal M}=0 ~~\rightarrow ~~\delta S_\text{G}|_{\partial \cal M}=0\,.
\ee
The variation $\delta S_\text{G}=0$ gives Einstein's equations, while the coordinates vary freely at the boundary. In contrast to the action $S_\text{EH}+S_\text{GHY}$ no unphysical boundary conditions for the lapse and the shift are required, indeed no boundary conditions for unphysical variables are allowed. This implies that the Hamiltonian of the covariant theory vanishes and no boundary term can change this. To change the Hamiltonian we might add a term $\partial_r (\ms N_t E)$ to the Lagrangian density, but this term contributes a non-vanishing boundary term $\propto \delta \ms N_t\ne 0$ to the variation of the action, and renders the variational problem ill-posed.

Let us briefly return to the concern formulated above that a path integral with fixed induced metric at $\partial \cal M$ might sum over classically distinct states, and give an over-counting of the microstates of a system. Consider the thermal partition function $Z$ as a path integral over Euclidean de Sitter space with Hubble parameter $H$. In the semiclassical limit and at vanishing energy the entropy equals the Euclidean action, which depends on the choice of boundary conditions 
\bea\label{partfunc}
\log(Z)=S(\text{EdS})=\begin{cases}S_\text{EH}+S_\text{GHY}={\pi\over G H^2}~~&\text{for}~~\delta\gamma_{\mu\nu}|_{\partial \cal M}=0\,,\\\hspace{47.1pt}S_\text{G}=0~~&\text{for}~~\delta{\ms  M}|_{\partial \cal M}=\delta{\ms  R}|_{\partial \cal M}=0\end{cases}\,.
\eea
The result (\ref{partfunc}) might give the impression that pure de Sitter space has a large number of microstates, $\sim e^{\pi/G H^2}$. However, pure de Sitter space has vanishing mass  $\ms M=0$, which can not be realized by the boundary conditions $\delta\gamma_{\mu\nu}=0$ unless coordinate transformations are prohibited. Correspondingly, the $\sim e^{\pi/G H^2}$ ``microstates'' include configurations with non-vanishing masses that are not pure de Sitter spacetimes. On the other hand, the covariant boundary conditions $\delta{\ms  M}|_{\partial \cal M}=\delta{\ms  R}|_{\partial \cal M}=0$ do not hold fixed the coordinates but instead restrict the spacetime to be pure de Sitter space. The gravitational action for de Sitter space vanishes, which implies that de Sitter space has one unique semiclassical microstate. This finding holds more generally: The Lagrangian density vanishes for all isotropic vacuum solutions with conserved covariant mass,
\be
{\cal L}_\text{G}|_{\dot{\ms M}=0}=0\,.
\ee
This is a main result of this paper\footnote{The reader shall not be confused by a misleading counter argument: Euclidean de Sitter space has no boundary, so one might be led to believe that derivative terms in the Lagrangian density do not contribute to the action $S_\text{EH}(+S_\text{GHY})=\pi/GH^2$. This is false, topological total derivative terms can change the classical action.
}.

This work is organized as follows. In \S\ref{setup} we discuss the setup of boundary conditions that do not depend on the coordinate choice and show that fixing the induced metric does not fix the mass of a Schwarzschild black hole. To build intuition and to connect to previous work we review the canonical action for ADM variables in \S\ref{wrongaction}. Finally, in \S\ref{correctaction} we change to covariant variables and derive a canonical action giving a well-posed variational problem, while the coordinate choice varies freely everywhere. We conclude with some comments in \S\ref{discuss}. We review the the variational principle for a non-relativistic particle in Appendix \ref{reviewbts}. Appendices \ref{app15}, \ref{app1}, \ref{app2} and \ref{app3} contain lengthy explicit expressions that would  distract from the main text.

\section{Setup and Boundary Conditions}\label{setup}
	\begin{figure}
\centering
\includegraphics[width=1\textwidth]{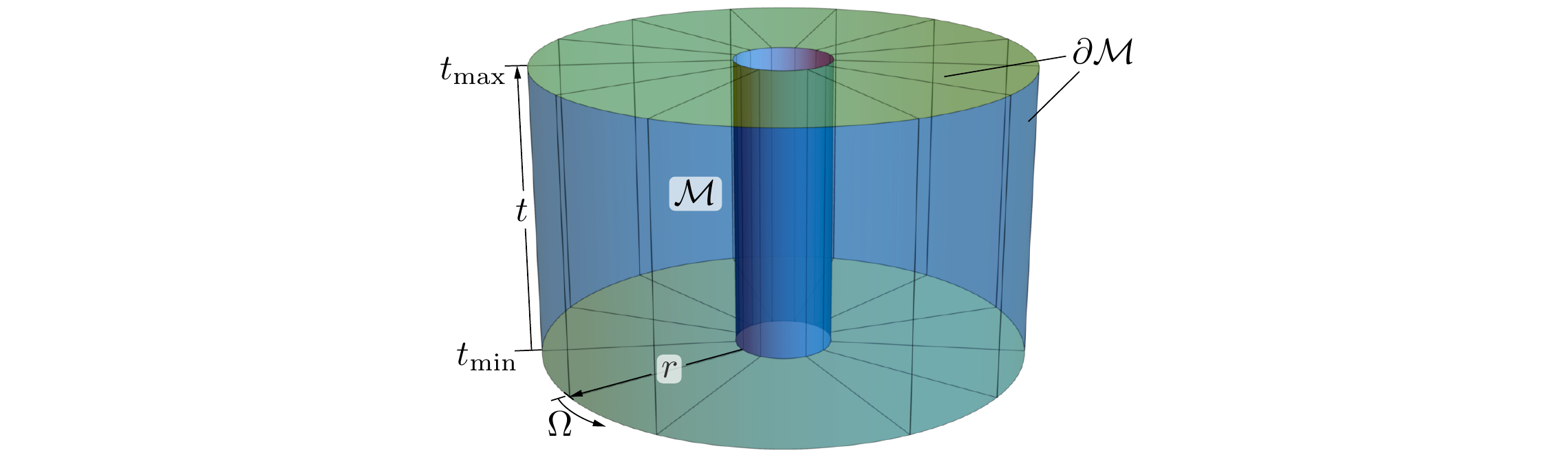}
\caption{\small Schematic illustration of the spacetime region ${\cal M}$, bounded by three-surfaces of constant $t$ and $r$ coordinates. $\Omega$ denotes two-sphere coordinates.}\label{spM}
\end{figure}

Consider an isotropic and bounded, but otherwise arbitrary spacetime-coordinate region $\cal M$ with homogeneous vacuum energy density. Without loss of generality we parametrize the general metric  with some isotropic coordinates $t$ and $r$, as well as angles $\Omega$ of the unit two-sphere. We choose the boundaries of ${\cal M}$ to be surfaces of constant coordinates,
\be
t_{\text{min}}\le t\le t_{\text{max}}\,,~~
r_{\text{min}}\le r\le r_{\text{max}}\,,
\ee
as illustrated in Figure \ref{spM}.

\begin{figure}
\centering
\includegraphics[width=1\textwidth]{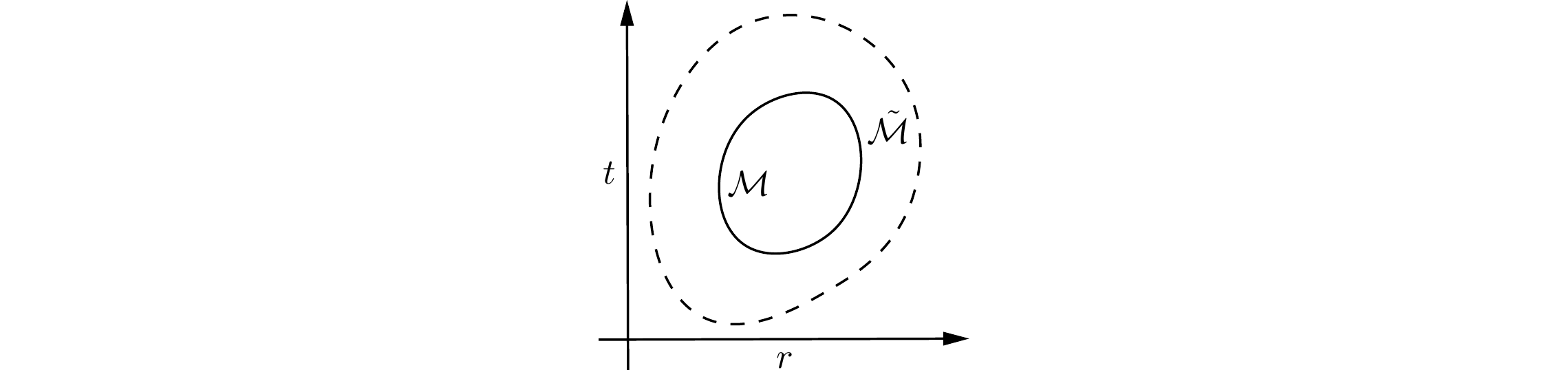}
\caption{\small Schematic illustration of two coordinate regions $\cal M$ and $\tilde{\cal M}$.}\label{cov}
\end{figure}

General covariance demands that the physical laws are independent of the coordinate region we study. Consider for example two coordinate manifolds $\cal M$ and $\tilde{\cal M}$, illustrated in Figure \ref{cov}. Let us formally denote the boundary conditions as ${\cal B}(t,r)$. The boundaries of the two coordinate regions are related by some coordinate transformation $\{t,r\}\rightarrow \{\tilde{t},\tilde{r}\}$, and therefore general covariance implies that the physical laws within $\cal M$ under boundary conditions imposed on the boundary $\partial{\cal M}$ are identical to the physical laws within $\cal M$ under boundary conditions imposed on the boundary $\partial\tilde{{\cal M}}$. This requirement of general covariance is satisfied if the boundary condition $\cal B$ transforms as a scalar under general coordinate transformations,
\be\label{covariant}
\{t,r\}\rightarrow \{\tilde{t},\tilde{r}\}\,:~~~{\cal B}(t,r)\rightarrow {\cal B}(\tilde{t},\tilde{r})\,.
\ee
We refer to the transformation property (\ref{covariant}) as (generally) covariant: the physical laws are invariant under coordinate transformations.

In this work we study general relativity and will therefore use the terms ``physical'' and ``(generally) covariant'' interchangeably, referring to configurations that are independent of  coordinate choices. Imposing spherical symmetry we can write the general metric in ADM form as
\be
\label{eq:NLRM}
ds^2 =  -{N_t^2}(t,r) \td t^2 + \Lambda^2(t,r) (\td r + {N_r} (t,r)\td t)^2 + R^2(t,r) \td \Omega_2^2 \,.
\ee
The lapse $N_t$ and the shift $N_r$ are arbitrary non-dynamical functions that set the gauge, while the variables $\Lambda$ and $R$ are dynamical.  Note that while $R(t,r)$ is covariant under coordinate re-parametrizations, $\Lambda(t,r)$ is not covariant and therefore will not be ideal to represent covariant boundary conditions. Throughout this work we denote covariant dynamical variables by Latin letters, while Greek letters denote non-covariant variables. Under a change of coordinates $r\rightarrow \tilde{r}(t,r)$ and $t\rightarrow \tilde{t}(t,r)$ we find the explicit transformation properties\footnote{The transformations of the lapse and the shift are given in (\ref{admtrafoos}).}
\bea\label{admreps}
R(t,r)&\rightarrow& \tilde{R}=R(\tilde{t},\tilde{r})\,,\\
\Lambda(t,r)&\rightarrow& \tilde{\Lambda}=\sqrt{\Lambda^2(\tilde{t},\tilde{r}) \left({\partial r\over  \partial{\tilde{r}}}+N_r(\tilde{t},\tilde{r}){\partial t\over  \partial{\tilde{r}}}\right)^2-\left(N_t(\tilde{t},\tilde{r}){\partial t\over \partial{\tilde{r}}}\right)^2}\,.\nonumber
\eea
Clearly, $\Lambda$ by itself contains no physical information and depends on the choice of coordinates. For example, we can choose coordinates such that $\Lambda=1$.

We seek a gravitational action with an action principle $\delta S_\text{G}=0$ that yields Einstein's field equations (\ref{einstein}). In order to solve for dynamics, and to check whether the variation of $S_\text{G}$ vanishes on the boundary ${\partial \cal M}$, we have to provide boundary conditions. In this work we demand that there exists a diffeomorphism that renders the metric on the boundary $g_{\mu\nu}|_{\partial \cal M}$ to be the Schwarzschild-(anti) de Sitter line element with mass  ${\ms  M}|_{{\partial \cal M}}$ and two-sphere radius ${\ms  R}|_{{\partial \cal M}}$,
\be
	\label{staticmetric}
		\td s^2|_{\partial \cal M} = -A(\ms  M,{\ms  R}) \td {\ms  T}^2 + A^{-1}({\ms  M},{\ms  R}) \td {\ms  R}^2 + {\ms  R}^2 \td\Omega_2^2\big|_{\partial {\cal M}} \,,
\ee
where $\ms T$ is the Killing time,  $\ms M$ is the mass, and we defined the convenient function
\be
A({\ms  M},{\ms  R})= 1 - \frac{2 G{\ms M}}{\ms R} - \frac{8\pi G\rho}{3} \ms R^2 \,.
\ee
A slight generalization of Birkhoff's theorem shows that the choice (\ref{staticmetric})  parametrizes the most general spherically symmetric metric with homogeneous energy density \cite{island,Schleich:2009uj}. Our covariant boundary conditions contain no information about the lapse and shift, but impose two Dirichlet constraints on the mass and radius,
\be\label{bcss}
\delta{\ms M}|_{\partial \cal M}=\delta \ms R|_{\partial \cal M}=0\,.
\ee
In order to connect with existing literature, and to explicitly construct an action  whose variation vanishes under the boundary conditions (\ref{bcss}), we can relate the variables appearing in (\ref{staticmetric}) to the ADM metric components \cite{Kuchar:1994zk}
\be\label{boundaryfixed}
\ms R(t,r)=R\,,~~\ms M(t,r)= { R\over 2G}\left(1-\frac{ R'^2}{\Lambda^2}+\frac{\left(\dot{R}-N_r R'\right)^2}{N_t^2}-\frac{8\pi G \rho}{3}   R^2\right),
\ee
where we suppressed the arguments on the right-hand sides. We will give the explicit derivation of (\ref{boundaryfixed}) in \S\ref{correctaction}.  The fields (\ref{boundaryfixed}) are covariant under smooth coordinate re-parametrizations. Using the transformation properties (\ref{admreps}) and (\ref{admtrafoos}) we simply find
\be\label{mrtrafos}
\{t,r\}\rightarrow \{\tilde{t},\tilde{r}\}\,:~~~\ms R(t,r)\rightarrow \ms R(\tilde{t},\tilde{r})\,,~~\ms M(t,r)\rightarrow \ms M(\tilde{t},\tilde{r})\,.
\ee
The second and third term in the expression for $\ms M$  conspire to give a non-trivial covariant variable. In \S\ref{correctaction} we will see that the particular choice in (\ref{boundaryfixed}) is convenient since then $-\ms M$ is a variable conjugate to the (non-local) Killing time $\int \td r {\ms T}'$ \cite{Kuchar:1994zk}.

We can express the variations of $\ms M$ and $\ms R$ in terms of the ADM variables
\bea
\delta \ms M&=&\frac{ {\pi_{\Lambda}} R'}{N_t}\delta N_r-\frac{ G {\pi_{\Lambda}^2}}{N_t R}\delta N_t+\frac{ R R'^2}{G {\Lambda}^3}\delta  \Lambda+\frac{ \ms M-{8 \pi\rho  R^3\over 3}  }{R}\delta R+\left( \frac{ N_r {\pi_{\Lambda}}}{ N_t}-\frac{ R R'}{G {\Lambda}^2 }\right)\delta R'-\frac{ {\pi_{\Lambda}}}{N_t}\delta{\dot{R}}\nonumber\\
\delta \ms R&=&\delta R\,,
\eea
where the momentum $\pi_\Lambda$ is given in (\ref{admmomenta}) below. Physical boundary conditions (\ref{bcss}) do not impose Dirichlet boundary conditions on the ADM variables $\Lambda$, $R$, $N_t$ and $N_r$, but instead fix the covariant mass $\ms M$. Conversely, imposing Dirichlet boundary conditions  on the ADM variables does not fix $\delta R'$ and $\delta \dot{R}$ and therefore the mass $\ms M$ is allowed to vary at the boundary.

\section{Canonical Action for  ADM Variables}\label{wrongaction}
As a warmup exercise, we now discuss how to construct an action principle for the ADM variables $\Lambda$, $R$, $N_{t}$ and $N_r$ that gives rise to Einstein's field equations\footnote{We are particularly thankful to Kate Eckerle and Ruben Monten for helpful discussions on this topic  \cite{bemtoappear}.}. 

In this section, let us consider the (non-covariant)  Dirichlet boundary conditions
\be\label{bcadm}
\delta  \Lambda|_{\partial \cal M}=\delta R|_{\partial \cal M}=\delta N_r|_{\partial \cal M}=\delta N_t|_{\partial \cal M}=0\,.
\ee
We will turn to more physical boundary conditions that are  covariant in the next section.

We integrate the Einstein-Hilbert action for the ADM metric (\ref{eq:NLRM}) over the angular coordinates and find a Lagrangian density ${\cal L}_\text{EH}$ that is linear in second and mixed derivatives of the fields \cite{bemtoappear}
\be
S_\text{EH}= {1\over 16\pi G}\int_{\cal M}d\Omega_2 \td t \td r\,\sqrt{g} ({\cal R}-16\pi G\rho)=\int_{t_\text{min}}^{t_\text{max}}{\td t\int_{r_\text{min}}^{r_\text{max}}{\td r ~ \mathcal{L}_{\text{EH}}(X, X', \dot{X}, X'', \dot{X}', \ddot{X})}}\,,
\ee
where  $X\in \{R\,,\Lambda\,,N_r\,,N_t\}$ can denote any of the ADM variables, and we give the fully explicit expression in (\ref{ehexplicit}). Varying the action and integrating any derivatives of field variations by parts allows us to write
\bea\label{ehvar}
\delta S_{\text{EH}}&=&\int_{r_{\text{min}}}^{r_{\text{max}}}\hspace{-2pt}{\td r \left[\sum_{I}\bigg({\partial \mathcal{L}_\text{EH}\over \partial \dot{X}_I}-  {{\partial_t}}{\partial \mathcal{L}_\text{EH}\over \partial \ddot X_I}-{\partial_r}{\partial \mathcal{L}_\text{EH}\over \partial \dot {X}'_I}\bigg)\delta X_I+ \pfrac{\mathcal{L}_\text{EH}}{\ddot X_I} \delta \dot{X}_I + \pfrac{\mathcal{L}_\text{EH}}{\dot{X}_I'} \delta X'_I \right]_{t_\text{min}}^{t_\text{max}} } \nonumber\\
&&+\int_{t_{\text{min}}}^{t_{\text{max}}}{\td t} \big[ ~~t ~\leftrightarrow ~r~~  \big]_{r_\text{min}}^{r_\text{max}}+\int_{\cal M}{\td t}{\td r}\sum_{I} \left(\text{~equations of motion~}\right)\delta X_I\,.\nonumber
\eea
Appendix \ref{app15} contains the explicit expression for the variation of a general action, including the equations of motion. We recognize the equations of motion as Einstein's field equations (\ref{einstein}). Unfortunately the derivatives of the variables are unfixed on $\partial {\cal M}$, so the variation of the action does not vanish under the boundary conditions (\ref{bcadm}) for solutions to the equations of motions. As expected, the Einstein-Hilbert action does not provide a well-posed variational principle under Dirichlet boundary conditions on the metric components.

In order to obtain a variational principle, we now bring the action into canonical form by adding total derivative terms to the Lagrangian density that do not change the equations of motion, but do change the boundary terms\footnote{This expression differs from \cite{runaways} and \cite{Fischler:1990pk,fmp2} by the boundary term $-2(N^t R)'$, but is the same choice as \cite{Kraus:1994by}.} \cite{Berger:1972pg,Kraus:1994by}
\bea\label{canactionadm}
{\mathcal L}_{\text{ADM}}&=&{\mathcal L}_{\text{EH}}+{\td {\cal F}^{\text{EH}}_r\over \td r}+{\td {\cal F}^{\text{EH}}_t\over \td t}\nonumber\\&=&{1\over 2 G}\bigg( 2{\dot{R}-N^r R^\prime \over N^t}\left[(N^r \Lambda R)^{\prime} -\partial_t (\Lambda R)\right]+{\Lambda\over N^t}(\dot{R}-N^r R^\prime)^2 \nonumber\\
&&~~~~~~\,-2\left(1-{R^\prime/ \Lambda}\right)(N^t R)^\prime + {N^t\over \Lambda}(\Lambda^2-R^{\prime2})-8 \pi G \Lambda N_t \rho  R^2 \bigg)\,,
\eea
where the functionals ${\cal F}^{\text{EH}}_{r,t}$ and more details are given in Appendix \ref{app2}. The Lagrangian density now  contains only first derivatives, so we can easily evaluate the momenta conjugate to the ADM variables
\bea\label{admmomenta}
			{\pi_{\Lambda}} = \frac{{N_r} R' - \dot{R}}{G {N_t}}R \,, ~~
			  \pi_R = \frac{({N_r} \Lambda R)' - \partial_t (\Lambda R)}{G {N_t} } \,,~~
		  \pi_{{N_t}} =0 \,,~~
		  \pi_{{N_r}} =0 \,.
\eea
Using the momenta (\ref{admmomenta}) we can cast the action in a simple canonical form 
\be\label{admactions}
S_{\text{ADM}}= \int{\td t \, \td r ~ \left( {\pi_{\Lambda}} \dot{L} + \pi_R \dot{R} - {N_t} {\cal H}_t^{\text{ADM}} - {N_r} {\cal H}_r^{\text{ADM}} \right)} - \int{\td t \left[ { L}_\text{BT}^\text{ADM} \right]_{r_\text{min}}^{r_{\text{max}}}}  \,,
\ee
where
\begin{align}
		\label{admhams}
			{\cal H}_t^{\text{ADM}} &= \frac{G \Lambda \pi_\Lambda^2}{2 R^2} - \frac{G}{R} {\pi_{\Lambda}} \pi_R + \frac1{2 G} \left[ \left( \frac{2 R R^\prime}{\Lambda} \right)^\prime - \frac{(R')^2}{\Lambda} - \Lambda \right] + 4\pi \Lambda R^2 \rho \,,  \nonumber\\
			{\cal H}_r^{\text{ADM}} &= R^\prime \pi_R - \Lambda \pi_{\Lambda}^\prime \,,  \nonumber\\
			{L}_\text{BT}^\text{ADM} &=   \frac{{N_t} R }{G}\left({R'\over \Lambda}-1\right) -{N_r} \Lambda {\pi_{\Lambda}}\,.
\end{align}
The lapse and the shift have vanishing conjugate momenta and are non-dynamical, while their equations of motion enforce the Hamiltonian constraints ${\cal H}_{t,r}^\text{ADM}=0$.

The contribution ${L}_\text{BT}$ from the spatial boundary deserves some comment. When evaluating it at large $R$ in an asymptotically flat spacetime, and in coordinates where $N_t=\sqrt{1-2 G M_{\text{ADM}}/R}\rightarrow 1$, $N_r=0$, $\Lambda=R'/N_t\rightarrow 1$, and $R=r$ we find ${L}_\text{BT}\rightarrow M_{\text{ADM}}$. As discussed in \cite{Arnowitt:1962hi,York:1972sj,Regge:1974zd,Gibbons:1976ue,Brown:1992br,Brown:1992bq} the inclusion of this term is not optional, it is required to eliminate the second derivatives from the Lagrangian density and render the variational principle well-posed given Dirichlet boundary conditions for the ADM variables. However, this boundary term clearly is not unique as we can add an arbitrary functional of the fields to $L_\text{BT}$ without affecting the variational problem. In our choice we included the term $-N_t R/G$ to prevent the boundary term from growing indefinitely at large radius. 

We stress that to obtain a variational principle it is not necessary to choose particular fall-off conditions (like $N_t\rightarrow 1$, $N_r\rightarrow 0$, etc.) or even to restrict to asymptotically flat spacetimes. To see this explicitly, consider the variation of the ADM action\footnote{Returning to the case of asymptotically flat spacetimes discussed above, we note that the variation of the action with respect to the lapse gives an energy, $-{\delta S_{\text{ADM}}/\delta N_t}={R(\Lambda-R')/ G\Lambda}\rightarrow M_{\text{ADM}}$.}
\bea\label{admaction}
&&\hspace{-4pt}\delta S_{\text{ADM}} =\int_{\cal M}{\td t}{\td r}\sum_{I} \left(\text{~equations of motion~}\right)\delta X_I-\int_{r_{\text{min}}}^{r_{\text{max}}}{\td r \left[\pi_R\delta R+{\pi_{\Lambda}} \delta \Lambda \right]_{t_\text{min}}^{t_\text{max}} } \nonumber\\
&&\hspace{-4pt}-\int_{t_{\text{min}}}^{t_{\text{max}}}{\td t} \left[{R(\Lambda-R')\over G\Lambda}\delta N_t+\Lambda{\pi_{\Lambda}} \delta N_r+N_r {\pi_{\Lambda}} \delta \Lambda+ \frac{\Lambda (G N_r \pi_R+N_t)-(N_t R)'}{G L}\delta R\right]_{r_\text{min}}^{r_\text{max}}\nonumber\,.\\
\eea
The boundary contributions contain the variations of all ADM variables at the boundary: $\delta \Lambda$, $\delta R$, $\delta N_t$ and $\delta N_r$. In contrast to the variation of the Einstein-Hilbert action (\ref{ehvar}), however, no variations of the derivatives appear. The boundary terms vanish under our Dirichlet boundary conditions (\ref{bcadm}), such that the variational principle $\delta S_{\text{ADM}} =0$ implies Einstein's field equations (\ref{einstein}). 

With the canonical action (\ref{admactions}) we succeeded in constructing an action with a well-defined variational principle that reproduces the desired equations of motion under Dirichlet boundary conditions. This is nice, but unfortunately does not solve the problem we are fundamentally interested in. We set out to find an action that not only has a well-defined variational principle under any boundary conditions, but under the covariant boundary conditions (\ref{physicalbc}). These boundary conditions are physically distinct,
\be
\delta \ms M|_{\partial \cal M}=\delta \ms R|_{\partial \cal M}=0~~~\centernot{\Longleftrightarrow}~~~\delta  \Lambda|_{\partial \cal M}=\delta R|_{\partial \cal M}=\delta N_r|_{\partial \cal M}=\delta N_t|_{\partial \cal M}=0\,.
\ee
The boundary terms in the variation of the canonical action $ S_{\text{ADM}}$ do not vanish when fixing the covariant mass $\ms M$ at the boundary, and therefore $ S_{\text{ADM}}$ does not have well-posed variational principle for solutions with fixed mass.  We already anticipated this in \S\ref{setup}. The ADM variable $\Lambda$ is a spatial scalar density and depends on the coordinate choice. Fixing $\Lambda$ at the boundary is meaningless unless we also provide information about the coordinates by fixing the lapse and the shift $N_t$ and $N_r$. The variables $\ms M$ and $\ms R$, on the other hand  fully specify the physical setup by themselves, regardless of the coordinate choice, which indeed remains entirely undetermined. The canonical action $ S_{\text{ADM}}$ would be suitable for a setup that involves some ``gauge fixing apparatus'' at the boundary, capable of fixing the coordinates and the variable $\Lambda$, but at present we have no evidence that such an apparatus can be constructed or that general covariance breaks down. In this work we stick to the principle of general covariance, and therefore cannot accept any boundary conditions for the spurious variables $N_t$ and $N_r$. We conclude that the canonical action (\ref{admaction}) does not provide a well-posed variational principle for physical systems, and $\delta S_{\text{ADM}}=0$ does not imply Einstein's equations.

\subsection{Addition of ADM boundary term}
Let us briefly comment on a commonly quoted attempt to cancel the non-vanishing boundary terms when we do not impose boundary conditions on the lapse and the shift. As above, we consider the asymptotic conditions $N_t=\sqrt{1-2 G M_{\text{ADM}}/R}\rightarrow 1$, $N_r=0$, $\Lambda=R'/N_t\rightarrow 1$, and $R=r$, such that the momentum vanishes $\pi_\Lambda\rightarrow 0$. From (\ref{admaction}) we see that the only non-vanishing boundary term with boundary conditions $\delta R=\delta \Lambda=0$ is
\be
\delta S_{\text{ADM}}\supset \int_{t_{\text{min}}}^{t_{\text{max}}}{\td t} \left[-{R(\Lambda-R')\over G\Lambda}\delta N_t\right]_{r_\text{min}}^{r_\text{max}}\,.
\ee
We can imagine adding a boundary term $N_t M_\text{ADM}$  to to the Lagrangian that is supposed to cancel the undesired boundary term for solutions of mass parameter $M_\text{ADM}$ \cite{Regge:1974zd,Kuchar:1994zk}. This is the prescription that Kucha\v{r} argues for in order to obtain a well-posed variational problem. Indeed, we find the variation
\be
\delta \left(S_\text{ADM}+\int\td t ~N_t M_\text{ADM}\right)\supset \int_{t_{\text{min}}}^{t_{\text{max}}}{\td t} \left[\left(M_\text{ADM}-{R(\Lambda-R')\over G\Lambda}\right)\delta N_t\right]_{r_\text{min}}^{r_\text{max}}\,.\label{admtry}
\ee
For one particular solution with mass $\ms M=M_\text{ADM}$ we have succeeded to cancel the boundary term. Unfortunately, Dirichlet boundary conditions on the ADM variables do not hold $\ms M$ fixed (a derivative $R'$ appears in (\ref{admtry}) that is undetermined by the boundary conditions), and so the boundary term does not vanish in general. Kucha\v{r} argues for a ``parametrization at infinities'', which treats the lapse at the boundary as a dynamical variable (this would be the clock of an observer), such that the equations of motion of motion enforce $\ms M=M_\text{ADM}$. While this approach does ensure vanishing boundary terms, it only works for one particular solution of gravity: The variational principle gives only the equations of motion for the mass $M_\text{ADM}$. If we are interested in a different solution we have to consider a new theory. We conclude that the ADM action, even with the addition of a boundary term $N_t M_\text{ADM}$, does not yield a well-posed variational principle that reproduces Einstein's equations in full generality (unless we fix the coordinates at the boundary).

\subsection{Example: de Sitter space}\label{admdsexample}
Let us briefly review the explicit example of (Euclidean) de Sitter space to illustrate the importance of boundary conditions for semiclassical and quantum phenomena.

\begin{figure}
\centering
\includegraphics[width=1\textwidth]{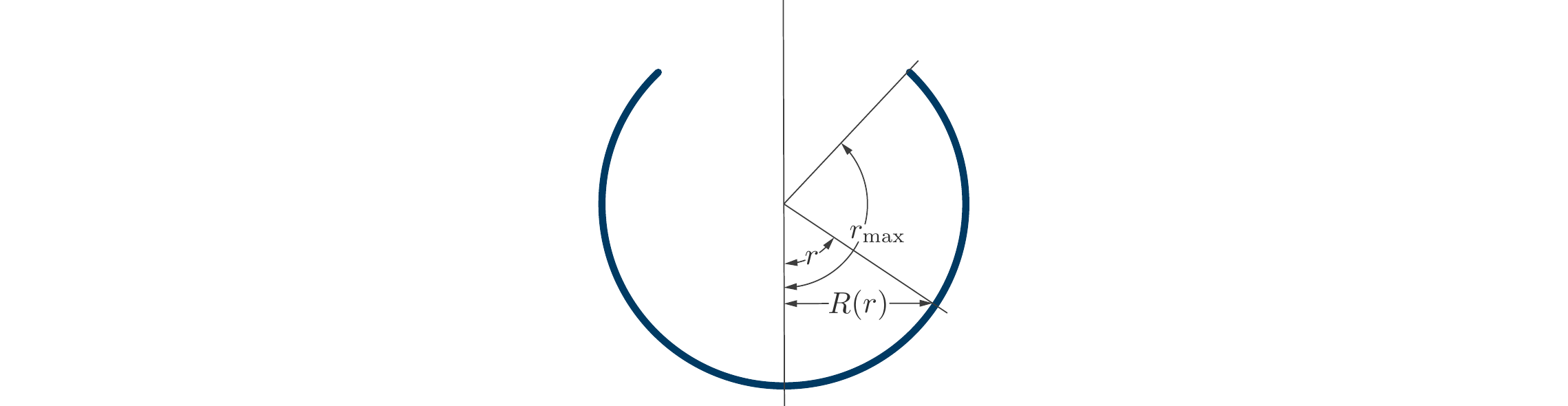}
\caption{\small Schematic illustration de Sitter space in closed slicing given in (\ref{asol}).}\label{dsfigure}
\end{figure}

To find a  solution for de Sitter spacetime we choose a simple gauge that identifies $\Lambda$ as the scale factor  $a(t)$ of an FRW cosmology,
\be\label{frwsols}
N_t(t,r)=1\,,~~N_r(t,r)=0\,,~~R(t,r)=a(t) \sin( r)\,,~~\Lambda(t,r)=a(t)\,.
\ee
This choice for $R$ solves the ${\cal H}_r=0$ constraint in (\ref{admhams}). Varying the action (\ref{admactions}) with respect to  $N_t$ and $\Lambda$ gives the remaining Hamiltonian constraint as well as the equations of motion for the scale factor, 
\be\label{eoma}
a\dot{a}^2=a(H^2 a^2-1)\,,~~2  a^3\ddot{a}=a^2(3 H^2 a^2-\dot{a}^2-1)\,,
\ee
where $H^2=8\pi G \rho/3$ is the Hubble constant. We can easily solve the constraint (\ref{eoma}) to find  solutions for ``nothing'' \cite{Vilenkin:1984wp,Vilenkin:1983xq} and de Sitter space in closed slicing
\bea\label{asol}
a(t)=\begin{cases}0~~~&\text{``nothing''}\\ H^{-1}\cosh(H t)~~~&\text{de Sitter space}\end{cases}\,.
\eea
The metric (\ref{eq:NLRMetricintro}) becomes the familiar metric for a closed FRW universe
\be
ds^2=-{ \td t^2}+a^2(t) d\Omega_3^2\,.
\ee
Imposing the constraint (\ref{asol}) and integrating over $r$, the remaining Lagrangian is given by
\be
L_{\text{ADM}}=\int_{r_\text{min}=0}^{r_\text{max}}\td r\,{\cal L}_\text{ADM}=\int_{r_\text{min}=0}^{r_\text{max}}\td r\,{a\over G}\left[{2 \sin ^2\left(\frac{r}{2}\right) (\cos (r)+2)-3 H^2 a^2 \sin ^2(r)}\right]\,.
\ee
Note that we were careful not to rush to pick our boundary $\partial {\cal M}$ to include an entire global slice of de Sitter space, which in our gauge would correspond to $r_\text{max}=\pi$ (see Figure \ref{dsfigure}). The choice of the boundary coordinates is unobservable and therefore should not matter in a physical system.

We can easily evaluate the action of Euclidean de Sitter space by integrating the $t$-coordinate from $-i\pi/2<t<i \pi/2$, or equivalently,
\be
S_\text{ADM}({\text{EdS}})=2i\int^{H^{-1}}_{a=0}\td a\,{L_{\text{ADM}}\over \dot{a}}={1\over G H^2} (r_\text{max}-2\sin(r_\text{max})+\sin(2r_\text{max})/2)\,.\label{admedss}
\ee
With $r_\text{max}=\pi$ we recover the famous result $S_\text{ADM}({\text{EdS}})=\pi/GH^2$.

Taken by itself, it is not immediately obvious that the action of Euclidean de Sitter space has any physical relevance. However, $S_\text{ADM}({\text{EdS}})$ does appear in measurable, supposedly physical quantities when we try to use the canonical action in semiclassical  gravity. The ratio of vacuum transition rates $\Gamma$ between two closed spatial slices of de Sitter space ($r_\text{max}=\pi$) with respective Hubble parameters $H_{\cal A}$ and $H_{\cal B}$ satisfies the detailed-balance relation \cite{Lee:1987qc} (see also \cite{runaways})
\be\label{detailedbal}
{\Gamma|_{{\cal A}\rightarrow {\cal B}}\over \Gamma|_{{\cal B}\rightarrow {\cal A}}}=e^{{ S}_\text{ADM}({\text{EdS}},\,\cal B)-{ S}_\text{ADM}({\text{EdS}},\,\cal A)}\,,
\ee
which looks like we can identify de Sitter space with an ergodic thermal system of entropy given by its Euclidean action. Unfortunately, the relation (\ref{detailedbal}) is dependent on our choice of boundary $\partial {\cal M}$ via $r_\text{max}$,
\be
{\pi-4\over 2G H^2}\le S_\text{ADM}({\text{EdS}})\le {\pi\over G H^2}\,.
\ee
So long as $r_\text{max}>\pi/2$ an observer at $r=0$ has no access to the boundary $\partial {\cal M}$ since it is beyond their cosmological horizon. Still, the vacuum transition rate the observer measures appears to depend on the boundary location. Traditionally, the choice $r_\text{max}=\pi$ is made, but we stress that this is an ambiguous assumption that has observable consequences and represents a manifestation of the measure problem. Similarly, the probability to tunnel  from ``nothing'' to de Sitter space in (\ref{asol}) is boundary dependent. Vilenkin argued for this probability to nucleate a universe from nothing to be given by \cite{Vilenkin:1984wp,Vilenkin:1983xq}
\be\label{edss1}
P\sim e^{-S_\text{ADM}({\text{EdS}})}\,.
\ee
Again, this result depends on the location of the boundary. 

To summarize, we see that  observables like vacuum decay rates or tunneling probabilities depend on the location of the (in principle arbitrary) boundary $\partial {\cal M}$. This was anticipated. In setting up the action we imposed Dirichlet boundary on all variables, including the lapse and the shift. This fixes the coordinates at the boundary and breaks general covariance, so we should not expect (semiclassical) observables to be covariant either. The results (\ref{detailedbal}) and (\ref{edss1}) apply for a universe with some gauge-fixing machinery (perhaps with ``Firewalls'', or other mythological creatures \cite{Almheiri:2012rt}) located at the boundary $\partial {\cal M}$ or if the mass $\ms M$ is allowed to vary in  pure de Sitter space. In the absence of a gauge-fixing apparatus the coordinate dependence of observables indicates that we failed to set up the calculation correctly. In the next section we will remedy this problem, perform the same calculation in a manifestly  covariant manner and find observables that are independent of coordinate choices.

\section{Canonical Action for Covariant Variables}\label{correctaction}
In the previous section we reviewed the construction of a canonical action for the ADM variables. The action $S_\text{ADM}$ yields Einstein's equation from an action principle, but only if we brake general covariance by fixing (some) of the coordinates at the boundary $\partial {\cal M}$ and allow the mass to vary. We now turn to the construction of a canonical action in terms of  covariant variables that hold the mass fixed, and we will see that the corresponding action principle is well-posed and independent of the coordinate choice.

Recall from \S\ref{setup} the boundary conditions we wish to impose at $\partial {\cal M}$. We demand that there exist some lapse and shift functions $N_{t,r}$ that render the metric at the boundary to be the Schwarzschild-(anti) de Sitter line element with fixed mass $\delta\ms M=0$ and two-sphere radius $\delta\ms R=0$. This means the metric at the boundary can be written as in (\ref{staticmetric}) above\footnote{Again, we defined $A({\ms  M},{\ms  R})= 1 - {2 G{\ms M}/\ms R} - {8\pi G \rho} \ms R^2/3$\,.}
\bea
	\label{staticmetric2}
		\td s^2 &=& -A(\ms  M,{\ms  R}) \td {\ms  T}^2 + A^{-1}({\ms  M},{\ms  R}) \td {\ms  R}^2 + {\ms  R}^2 \td\Omega_2^2 \nonumber\\
		&=&-{N_t^2} \td t^2 + \Lambda^2 (\td r + {N_r} \td t)^2 + R^2 \td \Omega_2^2 \,,
\eea
where in the second line we compared to the metric in ADM variables.  Let us first determine the relations between the new (covariant) and old (ADM) variables. Writing $\ms T=\ms T(t,r)$ and $\ms R=\ms R(t,r)$ we evaluate the metric components in (\ref{staticmetric2}) and find \cite{Kuchar:1994zk}
\be\label{relatestatictogeneral}
\Lambda=\sqrt{A^{-1}\ms R'^2-A \ms T'^2}\,,~~ R=\ms R\,,~~
N_t={\dot{\ms T} \ms R'-\ms T' \dot{\ms R}\over \sqrt{A^{-1}\ms R'^2-A \ms T'^2}}\,,~~N_r={\ms R'\dot{\ms R} A^{-1}-\ms T'\dot{\ms T} A\over A^{-1}{\ms R}'^2-A \ms T'^2}\,\,.
\ee
We substitute the definitions of the ADM momenta (\ref{admmomenta}) to derive the useful expressions
\be\label{tpandsuch}
\ms T'=-{L{\pi_{\Lambda}}\over A R}\,,~~\pi_\Lambda^2={R^2\over G^2}\left({R'^2\over \Lambda^2}-A(\ms M,\ms R)\right)\,.
\ee
Remember that $R$ and ${\pi_{\Lambda}}$ are spatial scalars, while $\Lambda$ and $R'$ are scalar densities that depend on the coordinate choice. The mass $\ms M$ and the Killing time $\ms T$, on the other hand, are a  scalars that do not depend on the coordinate choice. Correspondingly, fixing $R$ and $\Lambda$ cannot possibly fix $\ms R$ and $\ms M$ if the coordinate choice is allowed to vary.

Following Kucha\v{r}, we would like to identify the negative Killing time gradient as the momentum conjugate to the mass,  $\pi_{\ms M}\equiv-\ms T'$. At the same time, we still require that the Hamiltonian constraint ${\cal H}_r=0$ generates spatial diffeomorphisms. Taken together these requirements determine the momenta conjugate to $\ms R$ and $\ms M$. We can  express the new variables and momenta in terms of the ADM variables as\footnote{The expression for $\pi_{\ms R}$ is rather lengthy in ADM variables, so we avoid it here. It is straightforward to obtain the explicit form for $\pi_{\ms R}(\pi_{\Lambda},\pi_R,\Lambda,R)$ from the remaining relations in (\ref{oldnewtrafo}).}
\bea\label{oldnewtrafo}
{\ms R}=R\,,~~~\ms M=\frac{G \pi_\Lambda^2}{2 R}+\frac{R}{2 G}\left(1-{R'^2\over \Lambda^2}\right)-\frac{1}{3} 4 \pi  \rho  R^3\,,\hspace{154.pt}\\
\pi_{\ms R}=\frac{\pi_R R'-\pi_{\Lambda}'\Lambda-\pi_{\ms M}\ms M'}{ \ms R'}\,,~~~\pi_{\ms M}=\frac{GR {\Lambda}^3 {\pi_{\Lambda}} }{R^2 R'^2-G^2 \Lambda^2 \pi_\Lambda^2}\,,~~~\ms N_t=-{N_t\over \Lambda}\,,~~~\,\ms N_r=N_r\,,\nonumber
\eea
where we defined new (arbitrary) non-dynamical fields $\ms N_t$ and $\ms N_r$ that will yield a convenient expression of the action. We already showed in (\ref{mrtrafos}) that the dynamical variables $\ms R$ and $\ms M$ transform covariantly. Let us demand that the Hamiltonian densities given in (\ref{admhams}) remain unaffected by the change of coordinates, i.e.
$N_{r,t} {\cal H}^{\text{ADM}}_{r,t}=\ms N_{r,t} {\cal H}^{\text{G}}_{r,t}$. This leads to the new (and much simpler) Hamiltonian densities
\be\label{gravham}
{\cal H}^{\text{G}}_r=\pi_{\ms R} \ms R'+\pi_{\ms M} \ms M'\,,~~{\cal H}^{\text{G}}_t=A \pi_{\ms M}\pi_{\ms R}+A^{-1}{\ms M'\ms R'}\,.
\ee
Finally, the we can express the gravitational action with suitable boundary contributions in terms of the covariant variables
\bea\label{gravaction}
S_\text{G}= \int_{\cal M} \td t \td r~{\cal L}_\text{G}&=&  \int_{\cal M} \td t \td r~{\cal L}_\text{ADM}+{\td {\cal F}^\text{ADM}_r\over \td r}+{\td {\cal F}^\text{ADM}_t\over \td t}\nonumber\\
&=& \int_{\cal M} \td t \td r~\frac{{\ms  M}' \left[\left(\ms N_r^2-\ms N_t^2\right) {\ms  R}'-\ms N_r \dot{{\ms  R}}\right]+\dot{{\ms  M}} \left[\dot{{\ms  R}}-N_r {\ms  R}'\right]}{A({\ms  M},{\ms  R}) \ms N_t}\,\nonumber\\
&=&\int_{\cal M}\td t  \td r ~ \pi_{\ms M} \dot{\ms M} + \pi_{\ms R} \dot{\ms R} - {\ms N_t} {\cal H}_t^{\text{G}} - {\ms N_r} {\cal H}_r^{\text{G}} \,,
\eea
where the boundary terms do not change the equations of motion and are explicitly given by
\bea\label{explicitbtcov}
2 G{\cal F}^\text{ADM}_r&=&{\left(N_r R R'-G N_t {\pi_{\Lambda}}\right)\log \left(\frac{R R'-G \Lambda {\pi_{\Lambda}}}{R R'+G \Lambda {\pi_{\Lambda}}}\right) }+{2}{N_t R (1-R'/\Lambda)}+2 G \Lambda N_r {\pi_{\Lambda}}\,,\nonumber\\
2 G{\cal F}^\text{ADM}_t &=&-{R R' \log \left(\frac{R R'-G \Lambda {\pi_{\Lambda}}}{R R'+G \Lambda {\pi_{\Lambda}}}\right)}-2 G \Lambda {\pi_{\Lambda}}\,.
\eea
The action (\ref{gravaction}) is the main result of this paper, and to our knowledge has not appeared in the literature\footnote{In \cite{Kuchar:1994zk} Kucha\v{r} obtains an action that contains  terms similar to (\ref{explicitbtcov}), but they include further boundary contributions to accommodate their Dirichlet (or natural) boundary conditions on the ADM variables. For comparison with the ADM action, we express $S_\text{G}$ in terms of ADM variables in Appendix \ref{app3}.}.  First, note that the Lagrangian density transforms nicely like a scalar density of weight one
\be
\{t,r\}\rightarrow \{\tilde{t},\tilde{r}\}\,:~~~{\cal L}_\text{G}\rightarrow \left({\partial t\over  \partial\tilde t} {\partial r\over  \partial\tilde r}-{\partial t\over  \partial\tilde r}{\partial r\over  \partial\tilde t}   \right){\cal L}_\text{G}\,,
\ee
where we used the transformation properties of the new lapse and shift functions in (\ref{covtrafoos}). Varying the action with respect to all variables yield the relations between the momenta and velocities, as well as the Hamiltonian constraints
\be\label{momentaandhamconst}
\pi_{\ms M}=\frac{\dot{\ms R}-\ms N_r \ms R'}{A \ms N_t}\,,~~\pi_{\ms R}=\frac{\dot{\ms M}-\ms N_r \ms M'}{A \ms N_t}\,,~~{\cal H}^{\text{G}}_r={\cal H}^{\text{G}}_t=0\,.
\ee
It is straightforward (but tedious) to check that the equations of motion are unaffected by the change of variables. Considering the Poisson brackets it is easy to verify that the total Hamiltonian generates $t$ translations, while the component ${\cal H}^{\text{G}}_r$ generates $r$ translations,
\bea
&\left\{\ms M, {\ms N_t} {\cal H}_t^{\text{G}} + {\ms N_r} {\cal H}_r^{\text{G}}\right\}=\dot{\ms M}\,,~~~~~~&\left\{\ms M,{\cal H}_r^{\text{G}}\right\}={\ms M}'\,,\nonumber\\
&\left\{\ms R, {\ms N_t} {\cal H}_t^{\text{G}} + {\ms N_r} {\cal H}_r^{\text{G}}\right\}=\dot{\ms R}\,,~~~~~~~\hspace{1pt}&\left\{\ms R,  {\cal H}_r^{\text{G}}\right\}=\ms R'\,.
\eea

As desired, the new canonical action constructed above contains at most first derivatives of the variables. In contrast to the canonical ADM action (\ref{admactions}) however, ${\cal L}_\text{G}$ contains no derivatives of the non-dynamical variables $\ms N_{t,r}$. This means that no boundary terms containing $\delta N_{t,r}$ will appear in the variation of the action. Explicitly, we have
\bea\label{gravactionvar}
\delta S_{\text{G}} &=&\int_{\cal M}{\td t}{\td r}\sum_{I} \left(\text{~equations of motion~}\right)\delta \ms X_I+\int_{r_{\text{min}}}^{r_{\text{max}}}{\td r \left[\pi_{\ms R}\delta \ms R+\pi_{\ms M} \delta \ms M \right]_{t_\text{min}}^{t_\text{max}} }\nonumber \\
&&-\int_{t_{\text{min}}}^{t_{\text{max}}}{\td t \left[{\ms N_t\ms M' +\ms N_r A \pi_{\ms R} \over A}\delta \ms R+{\ms N_t\ms R' +\ms N_r A\pi_{\ms M} \over A}\delta \ms M \right]_{r_\text{min}}^{r_\text{max}} } \,,
\eea
where just as above $\ms X\in \{\ms R\,,\ms M\,,\ms N_r\,,\ms N_t\}$ can denote any of the variables. As expected, all boundary terms vanish under the two boundary conditions $\delta \ms M|_{\partial {\cal M}}=\delta \ms R|_{\partial {\cal M}}=0$. No boundary conditions on the unphysical variables $\ms N_{t,r}$ are required for the  principle of stationary action $\delta S_\text{G}=0$ to yield the equations of motion.

\subsection{Effective action}
With (\ref{gravaction}) we found a canonical form of the gravitational action with a well-defined variational principle that requires no boundary conditions for unphysical degrees of freedom. We now proceed to evaluate the action for solutions to the equations of motion. Varying the action with respect to $\ms N_{t}$ and $\ms N_{r}$ yields the Hamiltonian constraints ${\cal H}^{\text{G}}_r={\cal H}^{\text{G}}_t=0$. With (\ref{gravham}) we can write these constraints as
\be
\pi_{\ms R}=\ms M'=0\,.
\ee
The gravitational action for solutions to the constraints therefore becomes
\be\label{gravonshell}
S_\text{G}|_{{\cal H}^{\text{G}}_{t,r}=0}=\int_{\cal M} \td r\,  \pi_{\ms M}\delta \ms M\,,
\ee
where we immediately see that the gravitational action evaluates to zero for solutions to the equations of motion with constant mass, $\dot{\ms M}=0$. 

We can obtain a more explicit expression for the classical action by evaluating the integral over $\delta {\ms M}$ in coordinates where $\Lambda=1$ (see (\ref{eq:NLRM}) for the explicit metric), such that $\Lambda$ does not vary along the integration contour. In these coordinates the physical configuration is represented by the function $R(t,r)$, and using (\ref{oldnewtrafo}) and (\ref{momentaandhamconst}) the momentum becomes
\be
\pi_{\ms M}=A^{-1}{\sqrt{R'^2-A }}\,.
\ee
We  integrate the momentum $\pi_{\ms M}$ to evaluate the action 
\be\label{onshellaction}
S_\text{G}|_{\Lambda=1,\,{\cal H}^{\text{G}}_{t,r}=0}=\int_{r_\text{min}}^{r_\text{max}} \td r\left[{RR'\over 2 G}\log \left({2R'\over A} \left\{R'+\sqrt{R'^2-A}\right\}-1\right)-{R\over  G}\sqrt{R'^2-A}\right]_{R(t_\text{min},r)}^{R(t_\text{max},r)}
\ee
For static configurations (or classical turning points) the momenta vanish, ${\pi_{\ms M}}=0$, giving $R'^2=A$. Substituting this into the expression for the gravitational action we find that the integrand vanishes,
\be\label{integrateee}
{\cal L}_\text{G}|_{\pi_\Lambda=0,\,{\cal H}^{\text{G}}_{t,r}=0}=0\,.
\ee
The gravitational action vanishes for arbitrary classical trajectories between turning points. The integrated action is independent of the coordinate choice, so  we dropped the constraint to $\Lambda=1$ in (\ref{integrateee}). Importantly, since the integrand vanishes, this result is independent of the boundary ${\partial {\cal M}}$, as expected for covariant boundary conditions.

 Fischler, Morgan and Polchinski obtained an expression similar to the  classical gravitational action $S_\text{G}$ in (\ref{onshellaction}) for the ADM action $S_\text{ADM}$ \cite{Fischler:1990pk,fmp2,Kraus:1994by}. While their result appears to  dependend on the integration contour in phase space\footnote{We thank Kate Eckerle, Ruben Monten and Frederik Denef for discussion on the argument presented in \cite{Fischler:1990pk,fmp2,Kraus:1994by}.}, they inspired us to evaluate the unambiguous result (\ref{onshellaction}).

\subsection{Example: de Sitter space}
Let us  return to the discussion of de Sitter space as a concrete example. The de Sitter solution is static, so we already know that we will find a vanishing gravitational action $S_\text{G}=0$. In order to compare to the previous example in \S\ref{admdsexample}, however, we now review this solution in the covariant coordinates in more detail.

Working in the same gauge as we did in the previous section, we have for an FRW cosmology with (\ref{frwsols}) and (\ref{oldnewtrafo}) the covariant variables
\be
\ms N_t(t,r)={1\over a}\,,~~\ms N_r(t,r)=0\,,~~\ms M(t,r)={a\over 2G}\left(1-H^2 a^2+\dot{a}^2\right) \sin( r)\,,~~\ms R(t,r)=a\sin(r)\,.
\ee
The Lagrangian density appearing in the canonical action (\ref{gravaction}) is now given by
\be\label{lagrangianas}
{\cal L}_\text{G}={a \sin^4(r) \dot{a}^2\over G}\frac{ 2 a \ddot{a}+\dot{a}^2-3 H^2 a(t)^2+1}{\sin ^2(r) \dot{a}^2-\cos ^2(r)}-{1\over a}{\cal H}^\text{G}_t\,,
\ee
while the Hamiltonian constraints ${\cal H}_{t,r}^\text{G}=0$ yield the equations of motion for the scale factor we already found above in (\ref{eoma}). Comparing with the equations of motion we see that, as expected, the Lagrangian density (\ref{lagrangianas}) vanishes for classical solutions,
\be
{\cal L}_\text{G}|_{{\cal H}_{t,r}^\text{G}=0}=0\,.
\ee
This means that the gravitational action for (Euclidean) de Sitter space vanishes when we impose Dirichlet boundary conditions on the physical variables, $S_\text{G}(\text{EdS})=0$. The action is independent of the location of the boundary.

As a simple consistency check, let us integrate the derivative terms relating the Lagrangian densities ${\cal L}_\text{G}$ and ${\cal L}_\text{ADM}$ in (\ref{gravaction}) for the example of Euclidean de Sitter space. We find
\be\label{btint}
2i\int_{a=0}^{H^{-1}}\td a \dot{a}^{-1}\int_{0}^{r_\text{max}}\td r~{\td {\cal F}^\text{ADM}_r\over \td r}+{\td {\cal F}^\text{ADM}_t\over \td t}=-{1\over G H^2} (r_\text{max}-2\sin(r_\text{max})+\sin(2r_\text{max})/2)\,,
\ee
where the contribution due to $\td {\cal F}^\text{ADM}_t$ integrates to zero. Comparing (\ref{btint}) to the ADM action for the same setup in (\ref{admedss}), we see that the boundary terms we added precisely cancel the ADM action, giving a vanishing gravitational action $S_\text{G}$.

The observation that the gravitational action vanishes has far reaching consequences for (semiclassical) quantum gravity. For example, in \S\ref{admdsexample} we mentioned that the thin-wall vacuum decay rate between two Hubble parameters $H_{\cal A}$ and $H_{\cal B}$ satisfies a detailed balance relationship (\ref{detailedbal}), with entropy equal to the horizon area in units of $4G$ \cite{Lee:1987qc}. Using covariant boundary conditions, however, we find a vanishing entropy and thus equal rates for true- and false-vacuum decays \cite{tbvacuumdecay},
\be
\Gamma|_{{\cal A}\rightarrow {\cal B}}=\Gamma|_{{\cal B}\rightarrow {\cal A}}\,.
\ee
In a thermodynamic analogy this implies that de Sitter space has  one unique state, as opposed to the $e^{\pi/GH^2}$ states suggested by using non-covariant boundary conditions.

\section{Discussion}\label{discuss}
We studied the action principle for isotropic general relativity, which only allows physically observable boundary conditions and preserves general covariance. The action for isotropic and stationary spacetime regions vanishes. Therefore, the semiclassical description of general relativity differs significantly from the commonly used gravitational theory that breaks general covariance, but holds fixed the (unobservable) induced metric on a boundary. 

The framework introduced in this paper represents a novel yet radically conservative approach to semiclassical quantum gravity that has obvious implications for several long-standing problems we will discuss in future work.

\section{Acknowledgments}
We thank Frederik Denef, Kate Eckerle, Eanna Flanagan, Raphael Flauger, Thomas Hartmann, Oliver Janssen, Austin Joyce, Matthew Kleban, Liam McAllister, Ruben Monten, Henry Tye and Erick Weinberg for useful discussions. This work was supported in part by DOE under grants no. DE-SC0011941 and DE-SC0009919 and by the Simons Foundation SFARI 560536.

\appendix
\section{Action Principle and Boundary Terms}\label{reviewbts}
In this appendix we briefly review the importance of boundary terms in the Lagrangian for the variational principle and semiclassical phenomena. Consider a non-relativistic particle of energy $E$ moving in one dimension. The action is given by
\be
S=\int_{t_\text{min}}^{t_\text{min}} \td t~L=\int_{t_\text{min}}^{t_\text{min}} \td t~\left({1\over 2 }{m\dot{q}^2}-V(q)+E-{{\td  {\cal F}(q,\dot{q})\over\td  t}}\right)\,,
\ee
where $q(t)$ is the position, dots denote time derivatives and ${\td {\cal F}}/\td t$ are some total derivative terms. All terms in the Lagrangian have real coefficients. We only work up to second time-derivatives, assume Dirichlet boundary conditions, $\delta q(t_\text{min})=\delta q(t_\text{max})=0$ and we added a term $E$ to the Lagrangian, such that the Hamiltonian vanishes. We denote the the classical turning points by $a$ and $b$, then for a tunneling trajectory (i.e. $V(q)>E$ for $a<q<b$) we have the boundary conditions
\be
q(t_\text{min})=a\,,~~q(t_\text{max})=b\,.
\ee
We are interested in finding the equations of motion, as well as the tunneling probability between the classical turning points $a$ and $b$.

First, let us try to find the equations of motion from an action principle $\delta S=0$. The variation of the action is given by
\bea
\delta S&=&\int \td t~ \left( \left\{- {{\partial V\over\partial q}} +{\partial \over\partial q} {{\td {\cal F}\over\td t}}\right\}\delta q +\left\{m \dot{q} +{\partial \over\partial {\dot{q}}} {{\td {\cal F}\over\td t}}\right\}\delta \dot{q}+ {\partial \over\partial {\ddot{q}}}  {{\td {\cal F}\over\td t}}\delta \ddot{q}\right)\nonumber\\
&=&\left[\left\{m \dot{q} +{\partial \over\partial {\dot{q}}} {{\td {\cal F}\over\td t}}-{\partial \over\partial {\ddot{q}}}  {{\td {\cal F}\over\td t}}\right\} \delta{q}+{\partial \over\partial {\ddot{q}}}  {{\td {\cal F}\over\td t}}\delta \dot{q}\right]_{t_\text{min}}^{t_\text{max}}\nonumber\\
&&+\int \td t~  \bigg\{\hspace{-7 pt}\underbrace{-m \ddot {q}- {{\partial V\over\partial q}}}_{\text{equations of motion}}\hspace{-7 pt} +\underbrace{{\partial \over\partial q} {{\td {\cal F}\over\td t}} - {\partial \over \partial t}{\partial \over\partial {\dot{q}}} {{\td {\cal F}\over\td t}} +{\partial^2\over \partial t^2}{\partial \over\partial {\ddot{q}}}  {{\td {\cal F}\over\td t}}}_{=\,0}\bigg\}\delta q\,. \label{aaa3}
\eea
The Euler-Lagrange equation is independent of the total derivative term,
\be
m\ddot{q}=-{\partial V\over \partial q}\,.
\ee
However, whether the Euler-Lagrange equation follows from the principle of stationary action, $\delta S=0$, does depend on the extra term. The first boundary term in (\ref{aaa3}) vanishes under our boundary conditions $\delta q=0$. The second term does not vanish if the total derivative term contains second derivatives,
\be
\delta S=0 ~~\rightarrow~~ m\ddot{q}=-{\partial V\over \partial q} ~~\text{only if~~} {\partial \over\partial {\ddot{q}}}  {{\td {\cal F}\over\td t}}=0\,.
\ee
We conclude that the principle of stationary action yields equations of motion only if the  boundary conditions ensure that all boundary terms vanish.

One might be tempted to ignore this problem. After all, the Euler-Lagrange equations remain unaffected even in the presence of a non-vanishing boundary term that demolishes the principle of stationary action: $\delta S\ne 0$. So long as all we do is solve the equations of motion, indeed no issue arises. However, the classical action is affected and the Hamilton-Jacobi formalism fails. This immediately leads to wrong results for semiclassical computations.

Let us attempt to evaluate the tunneling probability between the turning points $a$ and $b$, knowingly ignoring any non-vanishing boundary terms. We know that in the WKB approximation the wave-function evolves as $\psi\sim \exp\{{i\int_a^b dq\,p}\}=\exp\{{i\int_{t_\text{min}}^{t_\text{max}}\td L}\}$, where in the second equation we used the convenient choice for E that ensures that the Hamiltonian vanishes, i.e. $L=p\dot{q}-H=p\dot{q}$. This choice is similar to the discussion in gravity, where the Hamiltonian vanishes as well. Avoiding any details (and the WKB matching conditions that would determine the sign in the exponent), we schematically have for the transmission coefficient
\be
T\sim |\psi|^2\sim \left|\exp\left\{ {i[S(t_\text{max})-S(t_\text{min})]}\right\}\right|^2\,.
\ee
We find
\be
2i[S(t_\text{max})-S(t_\text{min})]=2i[{\cal F}({t_\text{max}})-{\cal F}({t_\text{min}})] +2\int \td q\sqrt{2m (V-E)}\,. 
\ee
If the variational problem is not well-posed, the total derivative term can easily change the value of the classical action and thus the tunneling probability. We have to be very careful to use good boundary conditions for any semiclassical computation we are interested in.

\section{Variation of Second Derivative Action}\label{app15}
For our setup the variation of a general action containing up to second derivatives is given by
\bea
\delta S=&&\int_{r_{\text{min}}}^{r_{\text{max}}}{\td r \left[\sum_{I}\bigg({\partial \mathcal{L}\over \partial \dot{X}_I}-  {{\partial_t}}{\partial \mathcal{L}\over \partial \ddot X_I}-{\partial_r}{\partial \mathcal{L}\over \partial \dot {X}'_I}\bigg)\delta X_I+ \pfrac{\mathcal{L}}{\ddot X_I} \delta \dot{X}_I + \pfrac{\mathcal{L}}{\dot{X}_I'} \delta X'_I \right]_{t_\text{min}}^{t_\text{max}} } +\nonumber\\
&+&\int_{t_{\text{min}}}^{t_{\text{max}}}{\td t} \Bigg[\sum_{I}\bigg({\partial \mathcal{L}\over \partial X'_I}-  {{\partial_r}}{\partial \mathcal{L}\over \partial  X_I''}-{\partial_t}{\partial \mathcal{L}\over \partial \dot {X}'_I}\bigg)\delta X_I+{\partial \mathcal{L}\over \partial \dot{X}'_I}\delta \dot{X}_I+{\partial \mathcal{L}\over \partial X''_I}\delta X_I'\bigg]_{r_\text{min}}^{r_\text{max}}\nonumber\\
&+&\int_{\cal M}{\td t}{\td r}\sum_{I} \bigg(\underbrace{{\partial \mathcal{L}\over \partial X_I}- {{\partial_t}}{\partial \mathcal{L}\over \partial \dot{X}_I} - {{\partial_r}}{\partial \mathcal{L}\over \partial {X}'_I} +{{\partial_t^2}}{\partial \mathcal{L}\over \partial \ddot{X}_I}+{{\partial_r^2}}{\partial \mathcal{L}\over \partial {X}''_I}+{\partial^2_{tr}} {\partial \mathcal{L}\over \partial \dot{{X}}'_I} }_{\text{equations of motion}}\bigg)\delta X_I\label{appehvariation} \,.\nonumber\\
\eea

\section{Transformation Properties}\label{app1}
Some simple algebra yields the $\{t,r\}~\rightarrow \{\tilde{t},\tilde{r}\}$ transformation properties of all ADM variables appearing in the metric (\ref{eq:NLRM})  as 
\bea\label{admtrafoos}
&&\hspace{-6pt}R(t,r)\rightarrow \tilde{R}=R(\tilde{t},\tilde{r})\,,\\
&&\hspace{-6pt}\Lambda(t,r)\rightarrow \tilde{\Lambda}=\sqrt{\Lambda^2(\tilde{t},\tilde{r}) \left({\partial r\over  \partial{\tilde{r}}}+N_r(\tilde{t},\tilde{r}){\partial t\over  \partial{\tilde{r}}}\right)^2-\left(N_t(\tilde{t},\tilde{r}){\partial t\over \partial{\tilde{r}}}\right)^2}\,,\nonumber\\
&&\hspace{-6pt}N_t(t,r)\rightarrow\tilde{N}_t=\tilde{\Lambda}^{-1}{{\Lambda(\tilde{t},\tilde{r})N_t(\tilde{t},\tilde{r})\left({\partial t\over  \partial\tilde t}{\partial r\over  \partial\tilde r}-{\partial r\over  \partial\tilde t}{\partial t\over  \partial\tilde r}\right) }}\,,\nonumber\\
&&\hspace{-6pt}N_r(t,r)\rightarrow\tilde{N}_r=\tilde{\Lambda}^{-2}\hspace{-2pt}\left[{\Lambda^2(\tilde{t},\tilde{r})  \left({\partial r\over  \partial\tilde t}+N_r(\tilde{t},\tilde{r}) {\partial t\over  \partial\tilde t}\right)\left({\partial r\over  \partial\tilde r}+N_r(\tilde{t},\tilde{r}) {\partial t\over  \partial\tilde r}\right)-N_t^2(\tilde{t},\tilde{r})   {\partial t\over  \partial\tilde t} {\partial t\over  \partial\tilde r} }\right]\hspace{-2pt}.\nonumber
\eea

Using (\ref{admtrafoos}) and the definitions of the covariant variables in (\ref{oldnewtrafo}) we then have the transformations
\bea\label{covtrafoos}
&&\hspace{-6pt}\ms R(t,r)\rightarrow \tilde{\ms R}=\ms R(\tilde{t},\tilde{r})\,,~~~\ms M(t,r)\rightarrow \tilde{\ms M}=\ms M(\tilde{t},\tilde{r})\,,\\
&&\hspace{-6pt}\ms N_t(t,r)\rightarrow\tilde{\ms N}_t= \frac{\ms N_t\left({\partial t\over  \partial\tilde t} {\partial r\over  \partial\tilde r}-{\partial t\over  \partial\tilde r}{\partial r\over  \partial\tilde t}\right)}{\left({\partial r\over  \partial\tilde r}+\ms N_r {\partial t\over  \partial\tilde r}\right)^2-\ms N_t^2 \left({\partial t\over  \partial\tilde r}\right)^2}\,, \nonumber\\
&&\hspace{-6pt}\ms N_r(t,r)\rightarrow\tilde{\ms N}_r=\frac{{\partial t\over  \partial\tilde t} \left(\left(\ms N_r^2-\ms N_t^2\right) {\partial t\over  \partial\tilde r}+\ms N_r {\partial r\over  \partial\tilde r}\right)+{\partial r\over  \partial\tilde t} \left({\partial r\over  \partial\tilde r}+\ms N_r {\partial t\over  \partial\tilde r}\right)}{\left({\partial r\over  \partial\tilde r}+\ms N_r {\partial t\over  \partial\tilde r}\right)^2-\ms N_t^2 \left({\partial t\over  \partial\tilde r}\right)^2} \,.\nonumber
\eea

\section{Derivation of Canonical Action for ADM Variables}\label{app2}
The Einstein-Hilbert action for the metric (\ref{eq:NLRMetricintro}) evaluates to
\bea\label{ehexplicit}
&&S_\text{EH} =\int_{\mathcal M}{\td^4 x~\sqrt{g} \left( \frac{R}{16 \pi G} - \rho \right)}=\int_{\cal M} \td t \td r~{\cal L}_\text{EH}\nonumber\\
&=&\int_{\cal M} \td t \td r~ {1\over 2G}\Bigg[\frac{{N_r}^2 R^2 \Lambda''}{{N_t}}-\frac{{N_r}^2 R^2 \Lambda' {N_t'}}{{N_t^2}}+\frac{2 {N_r}^2 R \Lambda' R'}{{N_t}}+\frac{3 {N_r} R^2 \Lambda' {N_r'}}{{N_t}}+\frac{{N_r} \dot{{N_t}} R^2 \Lambda'}{{N_t^2}}\nonumber\\
&-&\frac{2 {N_r} R \dot{R} \Lambda'}{{N_t}}+\frac{R^2 \Lambda' {N_t'}}{\Lambda^2}+\frac{2 {N_t} R \Lambda' R'}{\Lambda^2}-\frac{2 \Lambda {N_r}^2 R {N_t'} R'}{{N_t^2}}+\frac{2 \Lambda {N_r}^2 R R''}{{N_t}}+\frac{\Lambda {N_r} R^2 {N_r'}'}{{N_t}}\nonumber\\
&+&\frac{\Lambda {N_r}^2  R'^2}{{N_t}}+\frac{\Lambda \dot{{N_t}} R^2 N_r'}{{N_t^2}}-\frac{\Lambda {N_r} R^2 {N_r'} {N_t'}}{{N_t^2}}+\frac{\Lambda R^2 \left({N_r'}\right)^2}{{N_t}}-\frac{2 \dot{\Lambda} R^2 N_r'}{{N_t}}+\frac{4 \Lambda {N_r} R {N_r'} R'}{{N_t}}\nonumber\\
&-&\frac{2 \Lambda R \dot{R} {N_r'}}{{N_t}}+\frac{2 \Lambda {N_r} \dot{{N_t}} R R'}{{N_t^2}}+\frac{\dot{\Lambda} {N_r} R^2 {N_t'}}{{N_t^2}}+\frac{2 \Lambda {N_r} R \dot{R} {N_t'}}{{N_t^2}}-\frac{2 {N_r} R^2 \dot{\Lambda}'}{{N_t}}-\frac{\Lambda R^2 \dot{N}_r'}{{N_t}}\nonumber\\
&-&\frac{2 \dot{\Lambda} {N_r} R R'}{{N_t}}-\frac{2 \Lambda {N_r} \dot{R} R'}{{N_t}}-\frac{2 \Lambda \dot{{N_r}} R R'}{{N_t}}-\frac{4 \Lambda {N_r} R \dot{R}'}{{N_t}}-\frac{R^2 {N_t''}}{\Lambda }-\frac{\dot{\Lambda } \dot{{N_t}} R^2}{{N_t^2}}\nonumber\\
&-&\frac{2 \Lambda \dot{{N_t}} R \dot{R}}{{N_t^2}}-\frac{2 R {N_t'} R'}{\Lambda }+\frac{R^2 \ddot{\Lambda }}{{N_t}}-\frac{2 {N_t} R R''}{\Lambda }-\frac{{N_t}  R'^2}{\Lambda }+\frac{2 \Lambda R \ddot{R}}{{N_t}}\nonumber\\&+&\frac{\Lambda  \dot{R}^2}{{N_t}}+\frac{2 \dot{\Lambda } R \dot{R}}{{N_t}}-\frac{\dot{{N_r}} R^2 \Lambda'}{{N_t}}+\Lambda {N_t} \Bigg]-4 \pi  R^2   \Lambda {N_t} \rho\,.\label{ehwobt}
\eea

The ADM action is related to the Einstein-Hilbert action through the addition of boundary terms
\be
S_\text{ADM} =\int_{\cal M} \td t \td r~\left({\cal L}_\text{EH}+{\td {\cal F}^{\text{EH}}_r\over \td r}+{\td {\cal F}^{\text{EH}}_t\over \td t}\right)\,,
\ee
where
\bea
{\cal F}^{\text{EH}}_r&=&{  R \over 2G  \Lambda N_t^2}\big[\Lambda^2 \dot{N}_r N_t R-\Lambda ^2 N_r N_t R N_r'+4 \Lambda^2 N_r N_t \dot{R}-\Lambda^2 N_r \dot{N_t} R-\Lambda N_r^2 N_t R \Lambda'\nonumber\\&&~~~~~~~~~~\hspace{4pt}+2 N_t R' (N_t-\Lambda N_r) (\Lambda N_r+N_t)+2 \Lambda \dot{\Lambda} N_r N_t R-2 \Lambda N_t^3+N_t^2 R N_t'\big]\nonumber\\
{\cal F}^{\text{EH}}_t&=&{ R\over 2G  N_t^2}{ \left(\Lambda N_r R N_t'-\dot{\Lambda} N_t R-2 \Lambda N_t \dot{R}\right)}\,.\label{explicitbtadm}
\eea
Combining the above expressions the yields the canonical ADM action (\ref{canactionadm}).

We can find the boundary terms (\ref{explicitbtadm}) since they are related to a first-derivative form of the four-dimensional gravitational action,
\bea
{\td {\cal F}^{\text{EH}}_r\over \td r}+{\td {\cal F}^{\text{EH}}_t\over \td t}&=&\int_{S^2}\td \Omega_2{ \partial_\nu \left(\sqrt{g} g^{\mu\nu}\Gamma^\rho_{\mu\rho}\right)-\partial_\mu \left(\sqrt{g} g^{\nu\rho}\Gamma^\mu_{\nu\rho}\right)\over 16\pi G }+{ \Lambda N_t\over 2G}+\nonumber\\&&+\partial_t\left(\frac{  \Lambda R^2 N_r'}{4 GN_t}\right)-\partial_r\left({ R N_t\over G}+\frac{  \Lambda \dot{N}_r R^2}{4 G N_t}\right)\,.\label{funrelation}
\eea
The perhaps surprising non-derivative term in (\ref{funrelation}) arises because integration of the four-dimensional derivatives over the angular coordinates $\Omega_2$ yields a non-derivative term that we have to subtract. The terms in the second line of (\ref{funrelation}) eliminate derivatives of the lapse and the shift in the action and implement the background subtraction.

\section{Gravitational Action in Terms of ADM variables}\label{app3}
For completeness and in order to allow some comparison to ${\cal L}_\text{ADM}$ we now express the Lagrangian density ${\cal L}_\text{G}$ in terms of the ADM variables,
\bea
&&2 G \left({\Lambda}^2 \left(\dot{R}-{N_r} R'\right)^2-{N_t^2}  R'^2\right){\cal L}_\text{G}=\nonumber\\
&&\frac{{\Lambda}^3  R'^4 {N_r}^4}{{N_t}}+\frac{2 {\Lambda}^3 R  R'^2 R'' {N_r^4}}{{N_t}}-\frac{2 {\Lambda}^3 R {N_t'}  R'^3 {N_r^4}}{{N_t^2}}+\frac{2 {\Lambda}^3 R \dot{{N_t}}  R'^3 {N_r^3}}{{N_t^2}}+\frac{6 {\Lambda}^3 R \dot{R} {N_t'}  R'^2 {N_r^3}}{{N_t^2}}\nonumber\\ &&-\frac{4 {\Lambda}^3 \dot{R}  R'^3 {N_r}^3}{{N_t}}-\frac{4 {\Lambda}^3 R  R'^2 \dot{R}' {N_r}^3}{{N_t}}-\frac{4 {\Lambda}^3 R \dot{R} R'R'' {N_r}^3}{{N_t}}-2 {\Lambda} {N_t}  R'^4 {N_r}^2+2 {N_t} R {\Lambda}'  R'^3 {N_r}^2\nonumber\\&&
+2 {\Lambda} R {N_t'}  R'^3 {N_r}^2+\frac{6 {\Lambda}^3 \dot{R}^2  R'^2 {N_r}^2}{{N_t}}+{\Lambda}^3 {N_t}  R'^2 {N_r}^2-8 G {\Lambda}^3 {N_t} \pi  R^2 \rho   R'^2 {N_r}^2\nonumber\\&&+\frac{2 {\Lambda}^3 R \ddot{R}  R'^2 {N_r}^2}{{N_t}}+\frac{8 {\Lambda}^3 R \dot{R} R'\dot{R}' {N_r}^2}{{N_t}}+\frac{2 {\Lambda}^3 R \dot{R}^2 R'' {N_r}^2}{{N_t}}-4 {\Lambda} {N_t} R  R'^2 R'' {N_r}^2\nonumber\\&&-\frac{2 {\Lambda}^3 R \dot{{N_r}}  R'^3 {N_r}^2}{{N_t}}-\frac{4 {\Lambda}^3 R \dot{R} {N_r'}  R'^2 {N_r}^2}{{N_t}}-\frac{6 {\Lambda}^3 R \dot{{N_t}} \dot{R}  R'^2 {N_r}^2}{{N_t^2}}-\frac{6 {\Lambda}^3 R \dot{R}^2 {N_t'} R'{N_r}^2}{{N_t^2}}\nonumber\\&&-2 {N_t} R \dot{{\Lambda}}  R'^3 {N_r}+4 {\Lambda} {N_t} \dot{R}  R'^3 {N_r}-2 {\Lambda} {N_t} R {N_r'}  R'^3 {N_r}+\frac{4 {\Lambda}^3 R \dot{{N_r}} \dot{R}  R'^2 {N_r}}{{N_t}}\nonumber\\&&-2 {N_t} R \dot{R} {\Lambda}'  R'^2 {N_r}-4 {\Lambda} R \dot{R} {N_t'}  R'^2 {N_r}+\frac{2 {\Lambda}^3 R \dot{R}^3 {N_t'} {N_r}}{{N_t^2}}+\frac{6 {\Lambda}^3 R \dot{{N_t}} \dot{R}^2 R'{N_r}}{{N_t^2}}\nonumber\\&&-2 {\Lambda}^3 {N_t} \dot{R} R'{N_r}+16 G {\Lambda}^3 {N_t} \pi  R^2 \rho  \dot{R} R'{N_r}+\frac{2 {\Lambda}^3 R \dot{R}^2 {N_r'} R'{N_r}}{{N_t}}+4 {\Lambda} {N_t} R  R'^2 \dot{R}' {N_r}\nonumber\\&&+4 {\Lambda} {N_t} R \dot{R} R'R'' {N_r}-\frac{4 {\Lambda}^3 \dot{R}^3 R'{N_r}}{{N_t}}-\frac{4 {\Lambda}^3 R \dot{R} \ddot{R} R'{N_r}}{{N_t}}-\frac{4 {\Lambda}^3 R \dot{R}^2 \dot{R}' {N_r}}{{N_t}}+\frac{{\Lambda}^3 \dot{R}^4}{{N_t}}\nonumber\\&&+\frac{{N_t}^3  R'^4}{\Lambda}+{\Lambda}^3 {N_t} \dot{R}^2-8 G {\Lambda}^3 {N_t} \pi  R^2 \rho  \dot{R}^2-{\Lambda} {N_t}^3  R'^2-2 {\Lambda} {N_t} \dot{R}^2  R'^2\nonumber\\&&+8 G {\Lambda} {N_t}^3 \pi  R^2 \rho   R'^2+2 {N_t} R \dot{\Lambda} \dot{R}  R'^2+2 {\Lambda} {N_t} R \dot{R} {N_r'}  R'^2+\frac{2 {\Lambda}^3 R \dot{R}^2 \ddot{R}}{{N_t}}+2 {\Lambda} R \dot{R}^2 {N_t'} R'\nonumber\\&&-4 {\Lambda} {N_t} R \dot{R} R'\dot{R}'+\frac{2 {N_t}^3 R  R'^2 R''}{\Lambda}-\frac{2 {N_t}^3 R \Lambda'  R'^3}{{\Lambda}^2}-\frac{2 {\Lambda}^3 R \dot{{N_r}} \dot{R}^2 R'}{{N_t}}-\frac{2 {\Lambda}^3 R \dot{{N_t}} \dot{R}^3}{{N_t^2}}\,.
\eea
Clearly the ADM variables are an inconvenient choice to express an action principle with covariant boundary conditions.

\bibliographystyle{JHEP}
\newpage
\bibliography{bubblerefs}
\end{document}